\documentclass{article}

\def\nin{\noindent}

\usepackage{hyperref}
\usepackage{cleveref}
\usepackage[scaled=0.92]{helvet}        
\usepackage{enumitem}			
\usepackage{amsmath}
\usepackage{pstricks}           
\usepackage{graphicx}
\usepackage[hang,flushmargin]{footmisc}
\usepackage{url}
\usepackage{float}
\crefformat{footnote}{#2\footnotemark[#1]#3}
\graphicspath{%
    {converted_graphics/}
    {//.PSF/.Mac/Volumes/Max's UNIX Home area/Under work/AA_Papers_in_progress/Exact_permutation/}
}
\begin{document}
\title{Disease universe: Visualisation of population-wide disease-wide associations}
\maketitle

\vspace{-0.2cm}
\centerline{{\large Max Moldovan\footnote[1]{Australian Institute of Health Innovation, University of New South Wales, Sydney, Australia.  $^2$The APAC Sale Group, Singapore. $^3$Graduate Institute of Medical Informatics, College of Medical Science and Technology, Taipei Medical University, Taipei, Taiwan. $^4$Institute of Biomedical Informatics, National Yang Ming University, Taipei, Taiwan. Correspondence should be addressed to M.M. (max.moldovan@gmail.com)  and Y.-C.L. (jaak88@gmail.com).}, \: Ruslan Enikeev$^{2}$, \: Shabbir Syed-Abdul$^{3}$, \: Alex Nguyen$^{3,4}$, }}

\vspace{0.22cm}
\centerline{{\large Yo-Cheng Chang$^3$ \: and \: Yu-Chuan Li$^{3}$}}

\vspace{0.55cm}
\nin  {\it We apply a force-directed spring embedding graph layout approach to electronic health records in order to visualise population-wide associations between human disorders as presented in an individual biological organism. The introduced visualisation is implemented on the basis of the Google maps platform and can be found at} {\color{blue} \url{http://disease-map.net}}. {\it We argue that the suggested method of visualisation can both validate already known specifics of associations between disorders and identify novel never noticed association patterns. }
 
\vspace{0.35cm}
\nin {\it Key words}: systems biology; electronic health records; diseasomics; visualisation; graph layout

\vspace{0.35cm}



\noindent It is known that many human disorders are positively associated, accompanying each other due to various, often unknown, genetic, bio-pathological or common risk factors$^{1}$. There is also evidence that some disorders tend to be associated negatively, playing a preventative role against each other, or due to other hypothesised but not properly understood reasons$^{2,3}$. We use population-wide electronic health records data to visualise how human disorders are positioned against each other in a population with respect to an individual biological organism. By doing so, we attempt to execute a {\it systems biology} approach in order to reveal complex mechanisms underlying pathogenesis of human disorders. It is important to note that, due to specifics of electronic health records$^{4}$,  together with biological mechanisms the method may reflect certain aspects of a healthcare system. For example, closely related but distinct diagnoses are often recorded against the same medical condition. This would induce a positive association between disorders due to healthcare administration rather than biological reasons.

\vspace{0.15cm}
Observing a (sub)-population of size $N$, suppose that over a period $T$ there were $C_A$ individuals with at least one occurrence of disorder $A$, and $C_B$ individuals with at least one occurrence of disorder $B$. Further, $C_{AB}$ individuals presented with both disorders $A$ and $B$, each observed at least once over the same period. Then the information can be summarised by the following $2\times2$ table:

\vspace{-0.1in}
\begin{table} [htb]
\caption{Occurrence counts of $A$ and $B$ in population of size $N$.}
\begin{center}
\begin{small}
\vspace{-0.1in}
\begin{tabular}{c|cc|c}
\hline
& \multicolumn{2}{c|}{\underline{Disorder $A$}} & \\
Disorder $B$	& $A$ present & $A$ absent & Total \\ \hline
$B$ present &  $C_{AB}$ & $\cdot$  & $C_B$ \\
$B$ absent & $\cdot$ & $\cdot$ & $-$  \\ \hline
Total & $C_A$ & $-$ & $N$ \\
\hline
\end{tabular}
\label{tab:table1}
\end{small}
\end{center}
\end{table}

Table~\ref{tab:table1} is an example of a $2 \times 2$ table with fixed margins. Provided that individuals are affected independently of each other, it is common to assume that the number of co-occurrences $C_{AB}$ can be characterised by a non-central hypergeometric distribution$^{5,6}$. Further, the association between $A$ and $B$ can be objectively measured by the odds ratio $OR_{AB} \in [0, \infty)$. In case of no association between $A$ and $B$, $OR_{AB}$ is expected to take the value of one. Positive and negative associations between $A$ and $B$ would correspond to expected values of $OR_{AB}$ greater and less than 1, respectively. 

\vspace{0.15cm}
It should be noted that due to estimation, there is uncertainty about $OR$ values obtained from the data. Such an uncertainty is usually handled by reporting confidence intervals corresponding to $OR$ estimates. In the present version of method implementation, we intentionally avoided using confidence intervals, p-values or other statistical tools normally involved in hypothesis testing. We did so in order to reflect the empirical information contained in the data without any subjective interpretation that could otherwise be introduced through, for example, the choice of a significance level.

\vspace{0.15cm}
We estimated odds ratios $OR_{ij}$ for each possible pair of disorders $i$ and $j$ observed in a population. We further used the reversed expit transform of log-odds ratio estimates as a measure of Euclidian distances between pairs of nodes representing individual disorders in the force-directed spring embedding graph layout algorithm,$^{7-9}$ see {\bf Supplementary Materials} for details. 

\begin{figure}[h!]
\begin{center}
\includegraphics[width=15.2cm]{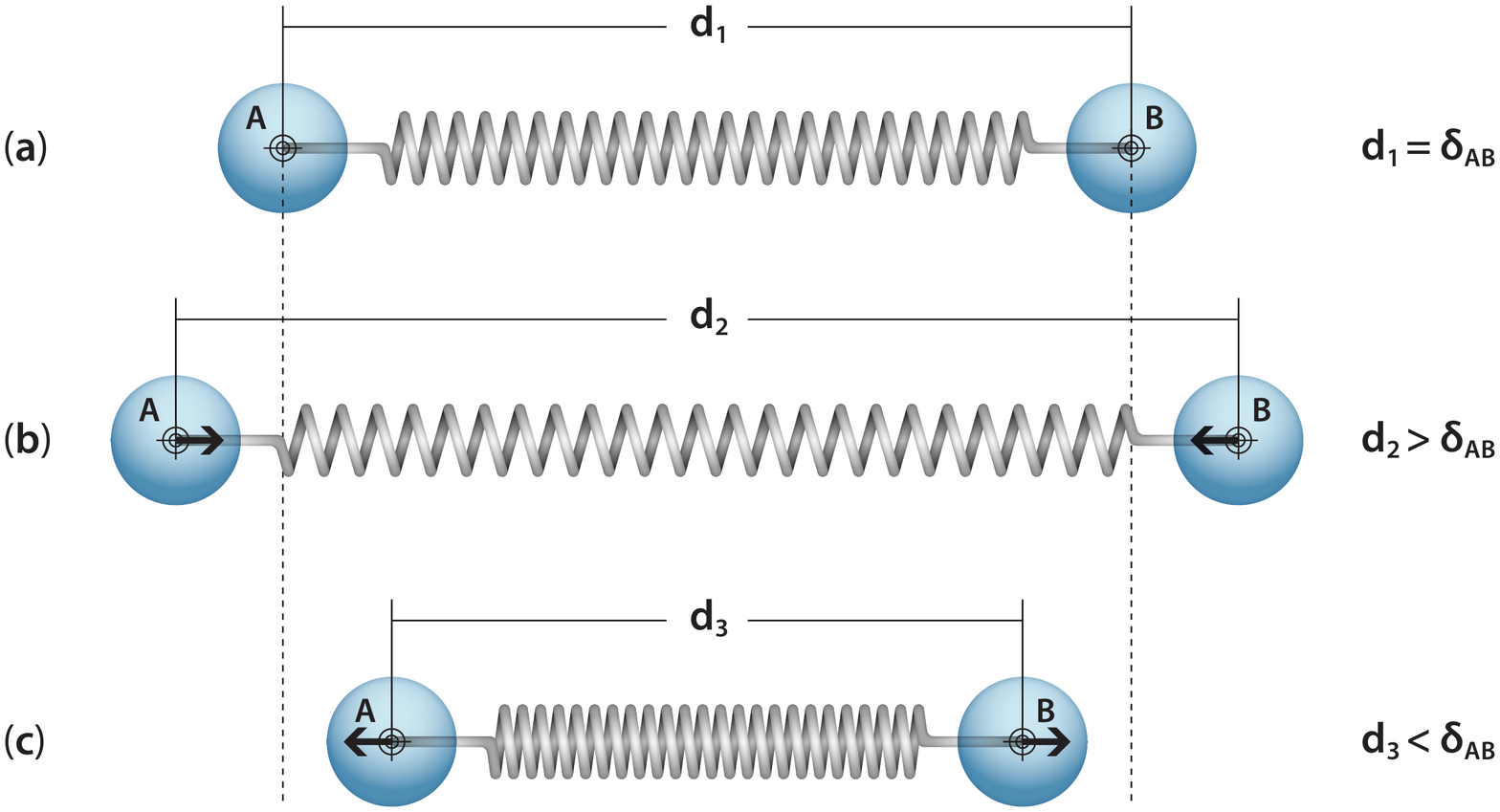}
\vspace{-1.0cm}
\caption{A single pair of nodes in three possible states. {\bf (a)} {\it Equilibrium}: the nodes neither attract nor repulse. {\bf (b)} {\it Attraction force}: the nodes are shifted far away from equilibrium and attempt to attract. {\bf (c)} {\it Repulsion force}: the nodes are closer than if they were in equilibrium and attempt to repulse.}
\label{fig:pic1}
\end{center}
\end{figure}

\vspace{0.15cm}
Imagine a single pair of nodes $A$ and $B$ positioned on a plane and connected by a spring of a certain {\it natural} length $\delta_{AB}$. When the distance between $A$ and $B$ is exactly $d_{AB} = \delta_{AB}$, the spring is in a state of equilibrium, creating neither attraction nor repulsion forces between the nodes (Figure~\ref{fig:pic1}a). Moving $A$ and $B$ far apart from each other would create an attraction force (Figure~\ref{fig:pic1}b), while moving $A$ and $B$ closer to each other would create a repulsion force between the nodes (Figure~\ref{fig:pic1}c). 

\begin{figure}[h!]
\begin{center}
\includegraphics[width=15.5cm]{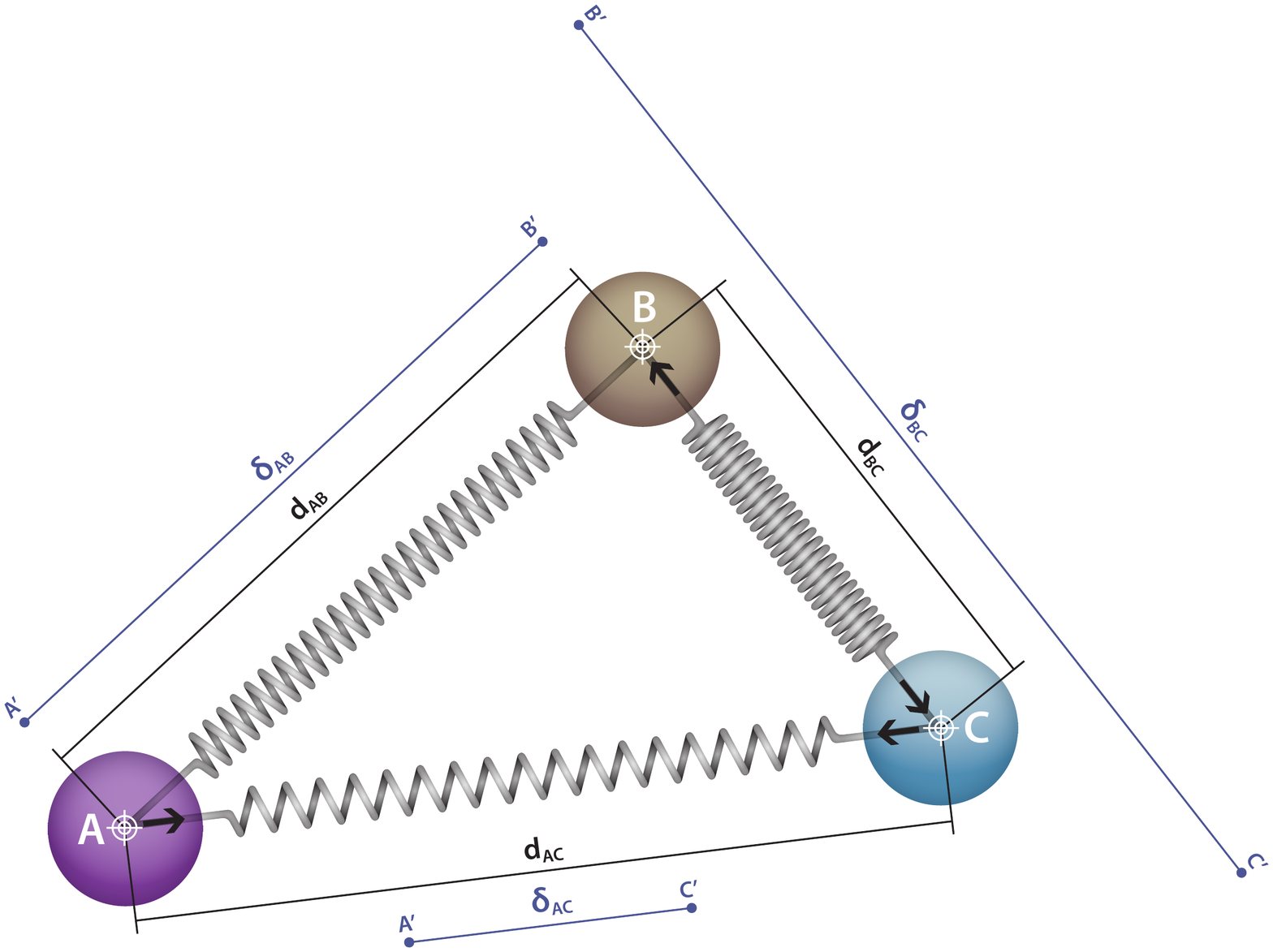}
\vspace{-0.50cm}
\caption{A hypothetical system of three nodes. The initial distances $\delta_{ij}$, $ij \in \{AB, AC, BC\}$ between the nodes are given by the theoretical lines $A'B'$, $B'C'$ and $A'C'$. The joint length of  $A'B' $ and $A'C'$  is less than the length of $B'C'$, i.e.  $\delta_{AB} + \delta_{AC} < \delta_{BC}$. As a result, for the nodes to connect, one or more of the initial distances between the pairs have to be distorted. The three possible states of springs are equilibrium (AB), attraction (AC) and repulsion (BC).}
\label{fig:pic2}
\end{center}
\end{figure}

Given the values of initial required distances $\delta_{ij}$ between multiple pairs of nodes, it is rarely possible to locate more than three nodes on a plain such that all required distances between them are satisfied exactly. In fact, it is not even always possible to locate three nodes keeping the pairwise distances in tact, see Figure~\ref{fig:pic2}. When distances between the nodes are not satisfied, springs connecting them are not in equilibrium, creating a certain force -- either attraction or repulsion. 

Aggregated forces created by out-of-equilibrium springs can be expressed by a specific function, drawn from the principle of physics (Hooke's law), leading to system's potential energy $U$. While a force can be positive (attraction) or negative (repulsion), an energy level is always non-negative irrespective of the sign of the force. Potential energy of a system of $M$ nodes connected by springs of varies stiffness, can be expressed as follows:

\begin{equation}
U = \frac{1}{2} \sum_{ij \in K} \left(  (d_{ij} - \widehat{\delta}_{ij})^2 \cdot \kappa_{ij} \right), \qquad K = \binom{M}{2}, \; i \not= j
\label{eq:full-energy1}
\end{equation}

\noindent where $d_{ij} \geq 0$ is an Euclidian distance between nodes $i$ and $j$, $\widehat{\delta}_{ij}$ is a natural length of a spring between nodes $i$ and $j$, $\kappa_{ij} > 0$ is an arbitrary parameter that defines the stiffness of a spring between $i$ and $j$, and $K = \binom{M}{2}$ is a number of all possible springs connecting $M$ nodes.

\vspace{0.15cm}
By varying pairwise Euclidian distances $d_{ij}$, the force-directed spring embedding graph layout algorithm performs a search for configuration of node locations such that system's potential energy $U$ is minimised. By finding the minimum energy $U$, we attempt to obtain a shape of a system of nodes in which competing forces largely compensate each other. Minimising function (\ref{eq:full-energy1}) is a complicated task due to the presence of multiple local minima, and it can rarely be guaranteed that the true global minimum is reached. However, we observed that most nodes have nearly constant ``designated" locations with respect to other nodes across alternative local minima achieved when minimising (\ref{eq:full-energy1}). 

\vspace{0.15cm}  
Electronic health records become an integral part of national healthcare systems worldwide, and it is essential to comprehensively utilise information contained in the growing number of databases. The method we present is one of the effective and informative tools for doing so. While the current realisation of the method has its obvious limitations, the presented maps are the first implementation of this kind and intended to set a reference benchmark for further developments in the same direction. We argue that the presented visualisation can both assist with validation of already know phenomena as well as with identification of novel, previously never noticed, association patterns related to functional aspects of medicine and healthcare$^{10}$. We suggest that the maps can be used for generating testable hypotheses and invite the reader to explore the vast amount of information contained in them at \hspace{0.001cm}  {\color{blue} \url{http://disease-map.net}}.

\vspace{1.2cm}
\noindent  {\bf ACKNOWLEDGMENTS}: We thank Hanna Kalkova and Andrey Stepanov for preparing Figures~\ref{fig:pic1} and \ref{fig:pic2}.

\vspace{0.35cm}
\noindent  {\bf FUNDING}: No formal funding was allocated towards the project.

\vspace{0.35cm}
\noindent  {\bf COMPETING FINANCIAL INTERESTS}: The authors declared no competing financial interests.

\vspace{0.9cm}
\nin{\large \bf References}

\begin{list}{}
{\leftmargin=1.0cm \itemindent=-1.0cm}

\item{} [1] Ferrannini, E. and Cushman, W.C. (2012) Diabetes and hypertension: the bad companions. {\it Lancet} {\bf 380}, 601-610.

\item{} [2] Rzhetsky, A, Wajngurt, D., Park, N. and Zheng, T. (2007) Probing genetic overlap among human phenotypes. {\it Proceedings of the National Academy of Sciences} {\bf 104}, 11694-11699.

\item{} [3]  Chou, F.H.-C., Tsai, K.-Y., Su, C.-Y. and Lee, C.-C. (2011) The incidence and relative risk factors for developing cancer among patients with schizophrenia: A nine-year follow-up study. {\it Schizophrenia Research} {\bf 129}, 97-103.

\item{}  [4] Hripcsak, G. and Albers, D.J. (2013) Next-generation phenotyping of electronic health records. {\it Journal of the American Medical Informatics Association} {\bf 20}, 117-121.

\item{}  [5] Lloyd, C.J. (1999) {\it Statistical Analysis of Categorical Data}. Wiley (New York).

\item{}  [6] Agresti, A. (2002) {\it Categorical Data Analysis}. 2d edition, John Wiley \& Sons.

\item{}  [7] Kamada, T. and Kawai, S. (1989) An algorithm for drawing general undirected graphs. {\it Information Processing Letters} {\bf 31}, 7-15.

\item{}  [8] Tunkelang, D. (1999) A numerical optimisation approach to general graph drawing. PhD thesis, Carnegie Mellon University: http://reports-archive.adm.cs.cmu.edu/anon/1998/CMU-CS-98-189.pdf

\item{}  [9] Hu, Y. (2005) Efficient, high-quality force-directed graph drawing. {\it The Mathematica Journal} {\bf 10}, 37-71. 

\item{}  [10] Syed-Abdul, S., Enikeev, R., Moldovan, M., Nguyen, A., Chang, Y.-C. and Li, Y.-C. (2013) Capturing and visualising human diseasomic associations. Manuscript.

\end{list}

\eject

\end{document}


\setcounter{page}{5}

\noindent {\bf SUPPLEMENTARY MATERIALS}

\begin{figure}[h!]
\begin{center}
\includegraphics[width=14cm]{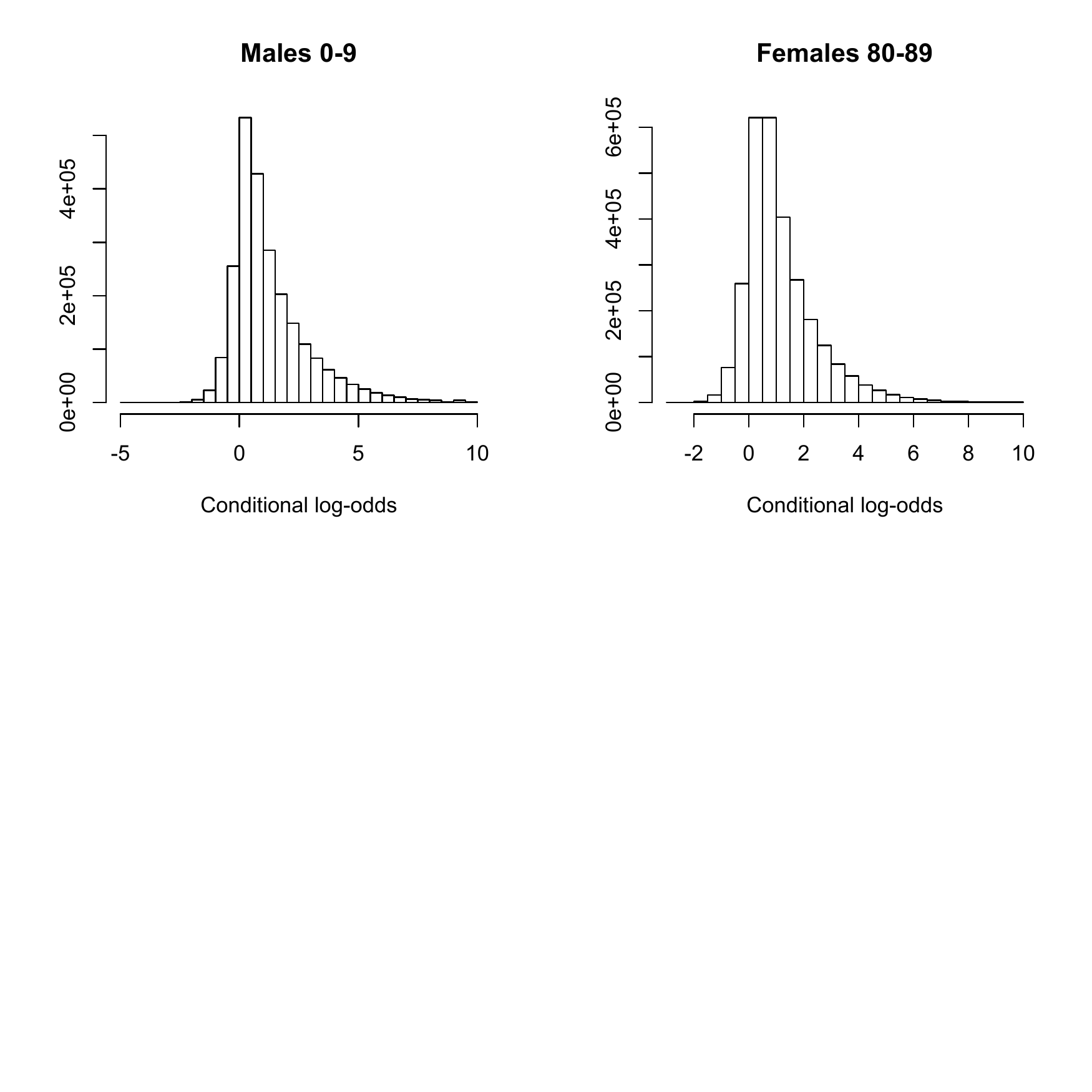}
\vspace{-6.5cm}
\caption{Empirical distribution of log-odds ratios for two distinct age-gender groups. Only the pairs with non-zero co-occurrences $C_{ij} > 0$ were considered. On the horizontal axis, zero is the point of no association, and larger values correspond to stronger associations, i.e. it is evident that human disorders tend to be positively associated. The same distribution pattern can be observed for other age-gender groups we considered.}
\label{fig:log-odds-hist}
\end{center}
\end{figure}

\vspace{0.35cm}
\noindent {\bf Distribution of disorder co-occurrences and the distance measure}. Let $C_{AB}$ be an outcome of a random variable $X$, and consider $B$ as a risk factor for $A$. Having the margins $C_A$ and $C_B$ in Table~\ref{tab:table2x2} fixed and assuming that subjects are affected independently of each other (which can be violated e.g. for infectious diseases), $X$ follows a non-central hypergeometric distribution $X \sim {\rm Hyper}(N,C_A,C_B)$ (Lloyd, 1999, p. 397; Agresti, 2002, p. 99):

\begin{equation}
\tag{i}
\Pr(X = C_{AB}) = \frac{ \binom{C_B} {C_{AB}} \binom{N - C_B}{C_A - C_{AB}}} {\binom{N}{C_A}} e^{\theta_{AB} C_{AB}}\\
\label{eq:hyper-geom}
\end{equation}

\noindent where $\binom{\cdot}{\cdot}$ is a binomial coefficient, $\max(0,C_A  + C_B - N) \leq {\bf C_{AB}} \leq \min(C_A,C_B)$,  $\theta \in (-\infty,+\infty)$ is a log-odds ratio, and $e = 2.718\ldots$ is the base of a natural logarithm. Conditional maximum likelihood estimates of $\theta$ were approximated by unconditional log-odds ratios:

\begin{equation}
\tag{ii}
\widehat{\theta}_{AB} = \ln \left(\frac{ C_{AB}(N - C_A + C_{AB} - C_B) } { (C_A - C_{AB})(-C_{AB} + C_B)} \right) = \ln \left(\frac{n_{11}n_{00}} {n_{01}n_{10}} \right)
\label{eq:uncodds}
\end{equation}

\noindent where $\ln(\cdot)$ is a natural logarithm. Switching the risk factor from being $B$ for $A$ to being $A$ for $B$ does not effect log-odds estimates. The empirical distribution of log-odds ratios is illustrated by Figure~\ref{fig:log-odds-hist}. Human disorders tend to be positively associated, and this pattern has already been documented, see Hidalgo {\it et al.} (2009).

Natural (equilibrium) lengths of springs between nodes $i$ and $j$ were obtained through the following {\it reversed} expit transform (Lloyd, 1999, p. 121):

\begin{equation}
\tag{iii}
\widehat{\delta}_{ij} = \frac {\exp(-\widehat{\theta}_{ij})}  {1+\exp(-\widehat{\theta}_{ij})}
\label{eq:nor}
\end{equation}

\noindent where $\widehat{\delta}_{ij} \in [0,1]$ by construction. Note that the sign on log-odds estimate $\widehat{\theta}$ was changed to the opposite (i.e. reversed), making stronger positive associations correspond to smaller values of $\widehat{\delta}_{ij}$. We do so in order make $\widehat{\delta}_{ij}$ to resemble Euclidian distances between the nodes.

\begin{table} [htb]
\caption{Records organised in a 2 $\times$ 2 table.}
\begin{center}
\begin{small}
\vspace{-0.16in}
\begin{tabular}{c|cc|c}
\hline
& \multicolumn{2}{c|}{\underline{Disorder $A$}} & \\
Disorder $B$	& $A$ present & $A$ absent & Total \\ \hline
$B$ present &  $C_{AB} \; [{\bf n_{11}}]$ & $C_B - C_{AB} \; {\bf [n_{01}]}$  & $C_B$ \\
$B$ absent & $C_A - C_{AB} \; {\bf [n_{10}]}$ & $C_{AB}+(N - C_B)-C_A \; {\bf [n_{00}]}$ & $N-C_B$ \\ \hline
Total & $C_A$ & $N-C_A$& $N$ \\
\hline
\end{tabular}
\label{tab:table2x2}
\end{small}
\end{center}
\end{table}

\vspace{0.35cm}
\noindent {\bf Force-directed spring embedding graph layout algorithm}. The current algorithm is a modified version of the algorithm underlying the internet map implementation at {\color{blue} \url{http://internet-map.net}}. Based on Hooke's law, the aggregate potential energy of a system is given by the following function:

\begin{equation}
\tag{iv}
U = \frac{1}{2} \sum_{ij \in K} \left(  (d_{ij} - \widehat{\delta}_{ij})^2 \cdot \kappa_{ij} \right), \qquad K = \binom{M}{2}, \; i \not= j
\label{eq:full-energy1}
\end{equation}

\noindent where $d_{ij} \geq 0$ is an Euclidian distance between nodes $i$ and $j$, $\widehat{\delta}_{ij}$ is a natural length of a spring between nodes $i$ and $j$, $\kappa_{ij}$ is an arbitrary parameter that defines the stiffness of a spring between $i$ and $j$ ($\kappa_{ij} = 1$ for all pairs in a special case with springs of equal stiffness) and $K = \binom{M}{2}$ is a number of all possible springs connecting $M$ nodes. The same can be rewritten in terms of coordinates:

$$U = \frac{1}{2} \sum_{ij \in K} \left(  \big(\sqrt{(X_i - X_j)^2 + (Y_i - Y_j)^2 } - \widehat{\delta}_{ij} \big)^2 \cdot \kappa_{ij} \right), \qquad K = \binom{M}{2}, \; i \not= j$$

\noindent where $(X_i, Y_i)$ and $(X_j, Y_j)$ are coordinates of nodes $i$ and $j$, respectively.  

\begin{figure}[h!]
\begin{center}
\includegraphics[width=14cm]{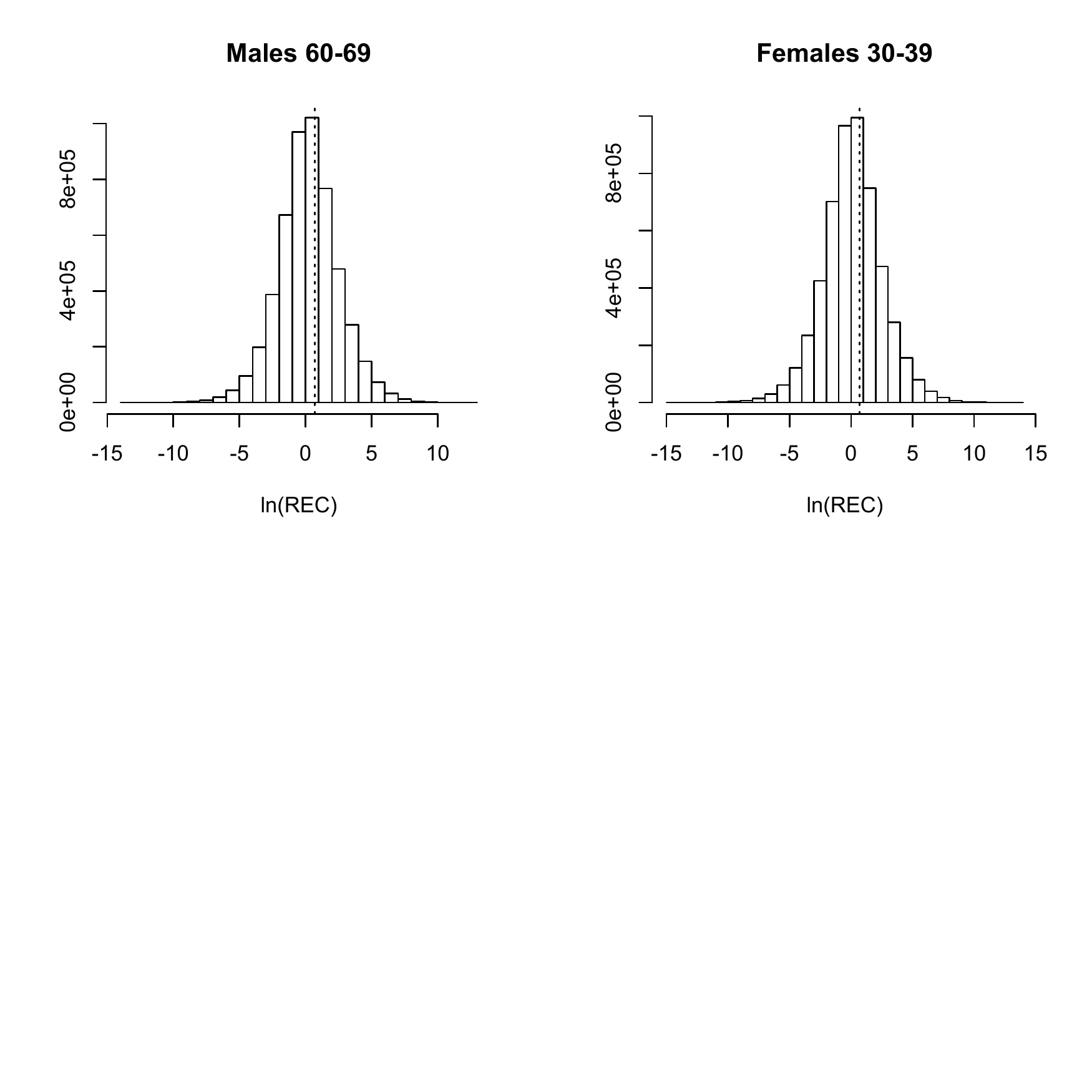}
\vspace{-6.5cm}
\caption{Empirical distribution of $\ln(REC_{ij})$ as given by (\ref{eq:rec}) for two distinct age-gender groups. Only the pairs with non-zero co-occurrences $C_{ij} > 0$ were considered. The imposed disease prevalence threshold corresponds to the horizontal line at $\ln(REC_{ij}) = \sqrt{2}$.  The pairs on the right of the horizontal threshold line were used for visualisations.}
\label{fig:log-rec-hist}
\end{center}
\end{figure}

\begin{figure}[h!]
\begin{center}
\includegraphics[width=12cm]{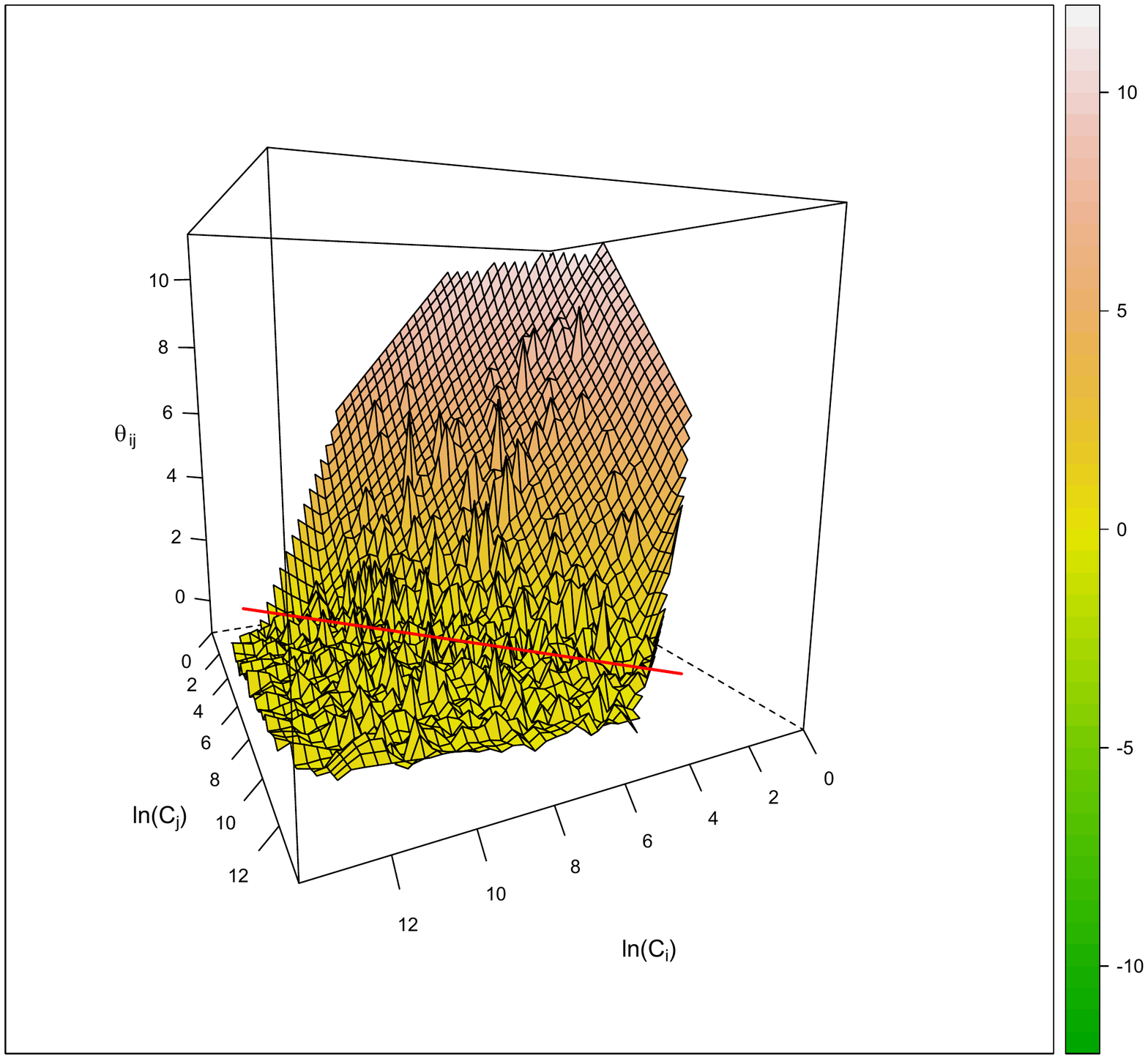}
\vspace{-2.5cm}
\caption{$\theta$-surface mesh plot as a function of $\ln(C_i)$ and $\ln(C_j)$. The line along the surface corresponds to values of $REC_{ij} = 1$, see equation (\ref{eq:rec}).  In the ``plateau" region in front of the line, $\theta$ estimates behave as they would be expected from a theoretical point of view, i.e. log-odds randomly vary around a certain horizontal expectation. For smaller counts $C_i$ and $C_j$,  $\theta$ estimates demonstrate an unexpected shift to the positive direction, i.e. upwards. Disorders corresponding this ``cliff" region were dismissed from the current method implementation by setting $C = \sqrt{2N}$ as a lower threshold on observed disorder counts.}
\label{fig:cliff}
\end{center}
\end{figure}

\vspace{0.35cm}
\noindent {\bf Cliff effect and the prevalence threshold}. Exploring the empirical distribution of log-odds ratios, we displayed a $\theta$-surface mesh plot as a function of $\ln(C_i)$ and $\ln(C_j)$ (Figure~\ref{fig:cliff}). This visualisation has revealed that log-odds estimates exhibit anomalous behaviour in the region of smaller counts $C_i$ and $C_j$. We have named this anomaly a {\it cliff effect} and attributed it to exceptionally high positive associations between certain pairs of disorders as observed in the context of the entire population and reflected by odds ratio estimates. In particular, the expected value of $X$ in the hypergeometric distribution function (\ref{eq:hyper-geom}) when $\theta = 0$, i.e there is no association between disorders, is given by (Agresti, 2002, p. 93):

\begin{equation}
\tag{v}
REC_{ij} = \frac{C_i C_j}{N}
\label{eq:rec}
\end{equation}

\noindent where $REC$ stands for Random Expected Co-occurrence. In Figure~\ref{fig:cliff}, the red line along $\theta$-surface corresponds to values of $REC_{ij} = 1$. When minimising the system's energy $U$ in (\ref{eq:full-energy1}), including $\theta$ estimates for pairs that lie in the ``cliff" region (i.e. behind the line) would bias the attention of an optimisation algorithm towards smaller prevalence diseases. We have executed the following ad hoc solution for dealing with the identified effect. Firstly, we imposed the threshold $C = \sqrt{2N}$ on disease occurrence counts. This guarantees that $REC_{ij} > 2$ for all possible $C_i$ and $C_j$, see Figure~\ref{fig:log-rec-hist}. The meaning behind this restriction is to ensure that only $\theta$ estimates from the ``plateau" region on the $\theta$-surface in Figure~\ref{fig:cliff} are used (i.e. estimates that are away from the ``cliff"). The cost is that we dismissed small prevalence disorders that never exceeded $REC_{ij} = 2$ in any of the age-gender groups. Secondly to the imposed lower limit on the observed occurrence counts, we set the stiffness parameter of a spring between pairs $i$ and $j$ to $\kappa_{ij} = \ln (REC_{ij})$. This modification makes sure that estimates corresponding to the area close to the ``cliff" region are given less importance when minimising the energy function (\ref{eq:full-energy1}). 

\vspace{0.35cm}
\noindent {\bf Energy minimisation method}. Finding a global minimum of (\ref{eq:full-energy1}) is a complicated task due to the presence of multiple local minima of this function. Different approaches of global minimisation can be applied, but it can be rarely known when and if the global minimum is reached, unless a minimum energy level is known in advance. Our current implementation of energy minimisation is to use multiple local searches with the conjugate gradient algorithm from random starting positions in order to obtain a {\it master} map -- the map that includes diseases across the entire spectrum of age groups and both genders. In each of multiple attempts, the nodes are dropped on the map with random positions $(X,Y)$, and the conjugate gradient algorithm runs searching for the closest local minimum of $U$ in (\ref{eq:full-energy1}). If the new local minimum is less than the best (smallest) minimum recorded over previous attempts, it becomes the new best minimum. The procedure is repeated until the best minimum stops changing even after a reasonably large (4000, in our implementation) number of random allocation attempts, see Algorithm~\ref{alt:alg1}. The computational complexity of the algorithm is $O(n^2)$. The computations were done on a purpose-built cluster. 

The obtained master map served as a collection of starting points for the age and gender stratified maps, see Algorithm~\ref{alt:alg2}. Minimising (\ref{eq:full-energy1}) from a single set of starting points leads to a local minimum that could almost always be further improved by applying the minimisation approach used for obtaining the master map. However, we still used minimisation from the single set of starting points in order to make the maps comparable across age groups and genders. Table~\ref{tab:table-S1} reports the achieved minimum energy levels using the ``partial" minimisation as per Algorithm~\ref{alt:alg2} compared to the ``full" minimisation implemented through Algorithm~\ref{alt:alg1}. 

\begin{algorithm}
\caption{Energy minimisation for the master map.}
\label{alt:alg1}
\begin{algorithmic} 

\REQUIRE $\gamma \leftarrow 0.01$ \hfill/* tolerance for the change in objective function (\ref{eq:full-energy1})
\REQUIRE $s \leftarrow 1$ \hfill /* initial step size
\REQUIRE $\tau \leftarrow 0.9$	\hfill/* step decrease rate
\REQUIRE $s_{min} \leftarrow 0.000001$	\hfill /* minimum step tolerance

\REQUIRE $\widehat{\delta}_{ij}$ for $K = \binom{M}{2}$ pairs, \; $i \not= j$ \hfill /* pairwise natural lengths given by (\ref{eq:nor})

\REQUIRE $U_{current} \leftarrow +Inf$ \hfill /* current energy level to be reduced
\REQUIRE $cc \leftarrow 0$ \hfill /* random positions attempts counter
\REQUIRE $cc_{max} \leftarrow 4000$ \hfill /* maximum number of attempts with no energy reduction

\WHILE{$(cc < cc_{max})$}

\STATE $(X_0,Y_0) \leftarrow random()$ \hfill /* drop nodes at random positions
\STATE $U_0 \leftarrow f_E(X_0,Y_0)$ \hfill /* value of objective function (\ref{eq:full-energy1})
\STATE $G \leftarrow \{-\nabla \left(f_E(X_0,Y_0)\right) \}$ \hfill /* define antigradients for the first step
\STATE $(\Delta X,\Delta Y)  \leftarrow  f_G(G)$ \hfill /* step direction
\STATE $(X,Y) \leftarrow (X_0,Y_0) + (\Delta X, \Delta Y) \cdot s$ \hfill /* current coordinates of nodes
\STATE $U \leftarrow f_E(X,Y)$ \hfill /*  current value of objective function (\ref{eq:full-energy1})
\STATE $\Delta U \leftarrow (U_0 - U)$  \hfill /* change in energy

\WHILE{$(\Delta U > \gamma) \; \& \; (s > s_{min}) $}

	\STATE $G_C \leftarrow \{\nabla_C \left(f_E(X_0,Y_0; X,Y)\right) \}$ \hfill /* evaluate conjugate gradients
	\STATE $(\Delta X,\Delta Y)  \leftarrow  f_{CG}(G_C)$ \hfill /* step direction
	\STATE $(X_{temp},Y_{temp}) \leftarrow (X,Y) + (\Delta X, \Delta Y) \cdot s$ \hfill /* trial coordinates of nodes
 	\STATE $U \leftarrow f_E(X_{temp},Y_{temp})$ \hfill /* current value of the objective function
	
		\IF{$U < U_0$}
	
			\STATE $\Delta U  \leftarrow (U_0 - U)$ \hfill  /* update change in energy
			\STATE $U_0 \leftarrow U$ \hfill	/* update preceding energy value
			\STATE $(X_0,Y_0) \leftarrow (X,Y)$ \hfill  /* update preceding coordinates
			\STATE $(X,Y) \leftarrow (X_{temp},Y_{temp})$ \hfill  /* assign the values of current coordinates

		\ELSE
			
			\STATE $s \leftarrow s \cdot \tau$ \hfill /* reduce step size	

		\ENDIF

\ENDWHILE

\IF{$U_0 < U_{current}$}

\STATE $U_{current} \leftarrow U_0$ \hfill /* update minimum energy value
\STATE $(X_{current}, Y_{current}) \leftarrow (X,Y)$ \hfill /* update coordinates
\STATE $cc \leftarrow 0$ \hfill  /* set attempts count to zero

\ELSE 

\STATE $cc \leftarrow cc + 1$ \hfill  /* next attempt

\ENDIF

\ENDWHILE

\STATE $(X_{master},Y_{master}) \leftarrow (X_{current}, Y_{current})$
\RETURN $(X_{master},Y_{master})$ \hfill /* nodes' coordinates under minimum energy achieved

\end{algorithmic}
\end{algorithm}

\begin{algorithm}
\caption{Energy minimisation for age and gender stratified maps.}
\label{alt:alg2}
\begin{algorithmic} 

\REQUIRE $\gamma \leftarrow 1e\mbox{-}5$ \hfill/* tolerance for the change in objective function (\ref{eq:full-energy1})
\REQUIRE $s \leftarrow 1$ \hfill /* initial step size
\REQUIRE $\tau \leftarrow 0.9$	\hfill/* step decrease rate
\REQUIRE $s_{min} \leftarrow 0.000001$	\hfill /* minimum step tolerance

\REQUIRE $\widehat{\delta}_{ij}$ for $K = \binom{M}{2}$ pairs, \; $i \not= j$ \hfill /* pairwise natural lengths given by (\ref{eq:nor})

\STATE $(X_0,Y_0) \leftarrow (X_{master},Y_{master})$ \hfill /* use coordinates from the master map as starting points
\STATE $U_0 \leftarrow f_E(X_0,Y_0)$ \hfill /* the value of objective function (\ref{eq:full-energy1})
\STATE $G \leftarrow \{-\nabla \left(f_E(X_0,Y_0)\right) \}$ \hfill /* define antigradients for the first step
\STATE $(\Delta X,\Delta Y)  \leftarrow  f_G(G)$ \hfill /* step direction
\STATE $(X,Y) \leftarrow (X_0,Y_0) + (\Delta X, \Delta Y) \cdot s$ \hfill /* current coordinates of nodes
\STATE $U \leftarrow f_E(X,Y)$ \hfill /*  current value of objective function (\ref{eq:full-energy1})
\STATE $\Delta U \leftarrow (U_0 - U)$  \hfill /* change in energy

\WHILE{$(\Delta U > \gamma) \; \& \; (s > s_{min}) $}

	\STATE $G_C \leftarrow \{\nabla_C \left(f_E(X_0,Y_0; X,Y)\right) \}$ \hfill /* evaluate conjugate gradients
	\STATE $(\Delta X,\Delta Y)  \leftarrow  f_{CG}(G_C)$ \hfill /* step direction
	\STATE $(X_{temp},Y_{temp}) \leftarrow (X,Y) + (\Delta X, \Delta Y) \cdot s$ \hfill /* trial coordinates of nodes
 	\STATE $U \leftarrow f_E(X_{temp},Y_{temp})$ \hfill /* current value of the objective function
	
		\IF{$U < U_0$}
	
			\STATE $\Delta U  \leftarrow (U_0 - U)$ \hfill  /* update change in energy
			\STATE $U_0 \leftarrow U$ \hfill	/* update preceding energy value
			\STATE $(X_0,Y_0) \leftarrow (X,Y)$ \hfill  /* update preceding coordinates
			\STATE $(X,Y) \leftarrow (X_{temp},Y_{temp})$ \hfill  /* assign values of current coordinates

		\ELSE
			
			\STATE $s \leftarrow s \cdot \tau$ \hfill /* reduce step size	

		\ENDIF

\ENDWHILE

\STATE $(X_{stratified},Y_{stratified}) \leftarrow (X,Y)$
\RETURN $(X_{stratified},Y_{stratified})$ \hfill /* nodes' coordinates under minimum energy achieved

\end{algorithmic}
\end{algorithm}

\begin{table}[htb]
\caption{Minimum achieved energy levels from partial and (attempted) full minimisation approaches.}
\begin{center}
\vspace{-0.16in}
\begin{tabular}{l | c c r r c} 
\hline
Group & Subjects followed ($N$) &	Disorder numbers &	\multicolumn{1}{c}{Partial} & \multicolumn{1}{c}{Full} &	Per cent improve \\
\hline
F 0-9 & 1,677,365 & 565 & 7,807.96 & 7,700.69 & 1.39 \\
F 10-19 & 1,595,057 & 743 & 9,470.09 & 9,166.34 & 3.31 \\
F 20-29 & 1,780,095 & 1041 & 22,897.04 & 22,268.39 & 2.82 \\
F 30-39 & 1,765,866 & 1136 & 25,914.10 & 25,387.60 & 2.07 \\
F 40-49 & 1,631,968 & 1243 & 31,126.22 & 30,913.37 & 0.69 \\
F 50-59 & 930,496 & 1251 & 33,451.35 & 33,334.80 & 0.35 \\
F 60-69 & 711,096 & 1271 & 36,129.76 & 36,056.87 & 0.20 \\
F 70-79 & 427,821 & 1177 & 29,935.15 & 29,857.36 & 0.26 \\
F 80-89 & 141,225 & 783 & 10,802.72 & 10,773.33 & 0.27 \\
F 90-99 & 8,532 & 176 & 318.26 & 315.74 & 0.80 \\
\hline
M 0-9 & 1,827,447 & 630 & 10,068.44 & 9,910.12 & 1.60 \\
M 10-19 & 1,678,415 & 721 & 9,451.03 & 9,346.16 & 1.12 \\
M 20-29 & 1,767,163 & 859 & 12,532.53 & 12,345.32 & 1.52 \\
M 30-39 & 1,737,715 & 948 & 14,263.07 & 14,099.18 & 1.16 \\
M 40-49 & 1,577,320 & 1090 & 19,485.22 & 19,454.02 & 0.16 \\
M 50-59 & 898,150 & 1065 & 20,296.22 & 20,247.20 & 0.24 \\
M 60-69 & 692,061 & 1163 & 26,737.58 & 26,563.97 & 0.65 \\
M 70-79 & 532,308 & 1225 & 30,740.90 & 30,622.10 & 0.39 \\
M 80-89 & 133,480 & 781 & 10,636.67 & 10,599.66 & 0.35 \\
M 90-99 & 4,769 & 151 & 240.25 & 238.70 & 0.65 \\
\hline
Master & 21,518,574 & 2298 & $-$ \quad\quad & 130,381.91 & $-$ \\
\hline
\end{tabular}
\label{tab:table-S1}
\end{center}
\end{table}

\vspace{0.35cm}
\noindent {\bf Results visualisation}. The Google maps platform ({\color{blue} \url{https://developers.google.com/maps/}}) was used to visualise the outcomes. The current implementation of the presented method can be found at {\color{blue} \url{http://disease-map.net}}. The sizes of the nodes are proportional to observed disease prevalence in the corresponding age-gender stratified sub-groups. The colour code corresponds to the broad disease categories (as per ICD9-CM classification) and is described in the `About' section on the website. All maps are shown in the same coordinate system with the same scale so they could be compared against each other.

\vspace{0.35cm}
\noindent {\bf Corrections for zero counts}. If any of the entries in Table~\ref{tab:table2x2} is zero, the log-odds ratio estimate $\widehat{\theta}_{AB}$ given by (\ref{eq:uncodds}) is undefined. Due to the lower prevalence threshold $C = \sqrt{2N}$ we imposed, the number of tables with zero counts is rather small, i.e. less than 0.5\% of all tables used for visualisation. We treat $C_{AB} = 0$ as evidence of negative association between $A$ and $B$ by setting $\widehat{\delta}_{AB} = 1$ in these cases. If any other cell in Table~\ref{tab:table2x2} happens equal zero, we apply the following correction:

$$\{C_{AB} +1; C_{A} +2; C_{B} +2; N+4\}  \equiv \{n_{11}+1; n_{10}+1; n_{01}+1; n_{00}+1\}.$$

\noindent This correction is equivalent to the Laplace estimator obtained by adding 1 to each cell in Table~\ref{tab:table2x2} (Greenland, 2000).

\vspace{0.35cm}
\noindent {\bf Data underlying the maps}. The records have been obtained from the Taiwanese national health insurance research database and cover the entire population of Taiwan over the period of three years (2000-2002). The same three-year observation window of the maximum available length has been used to record the counts corresponding to Table~\ref{tab:table2x2}. Disorder records are based on ICD9-CM (International Classification of Diseases, Ninth Revision, Clinical Modification), five-digit version. Each subject was noted over his of her first insurance claim starting from 01 January 2000, attributed to a certain age-gender group and followed for the rest of the period ending on 31 December 2002. Codes corresponding to $E$ and $V$ categories of ICD9-CM (external causes of injury and supplemental classification) were excluded from consideration. 

\vspace{0.35cm}
\noindent {\bf Intellectual property}. The presented method implementation has been registered as an invention through the University of New South Wales, see Moldovan {\it et al.} (2013).


\vspace{0.9cm}
\nin{\large \bf References}

\begin{list}{}
{\leftmargin=0.62cm \itemindent=-0.65cm}

\item{} Agresti, A. (2002) {\it Categorical Data Analysis}. 2d edition, John Wiley \& Sons.

\item{} Greenland, S. (1990) Small-sample bias and corrections for conditional maximum-likelihood odds-ratio estimators. {\it Biostatistics} {\bf 1}, 113-122. 

\item{} Hidalgo, C.A., Blumm. N., Barab\'{a}si, A.-L. and Christakis, N.A. (2009) A dynamic network approach for the study of human phenotypes. {\it PLoS Computational Biology} {\bf 5(4)}, e1000353.

\item{} Lloyd, C.J. (1999) {\it Statistical Analysis of Categorical Data}. Wiley (New York).

\item{} Moldovan, M., Enikeev, R., Syed-Abdul, S. and Li, Y.-C. (2013) Disease universe: Visualisation of population-wide disease-wide associations. Invention IPN-13-000062, NewSouth Innovations, Sydney, Australia, {\color{blue} \url{http://disease-map.net}}.

\end{list}

\eject

